\documentclass[
reprint,
nofootinbib,
 amsmath,amssymb,
aps,
]{revtex4-1}

\usepackage{graphicx}% Include figure files
\usepackage{dcolumn}% Align table columns on decimal point
\usepackage{bm}% bold math
\usepackage{hyperref}% add hypertext capabilities
\usepackage[mathlines]{lineno}% Enable numbering of text and display math
\usepackage{centernot}
\begin{document}

\preprint{APS/123-QED}

\title{Bag Formation from Gauge Condensate}

\author{Mahary Vasihoun}
\email{maharyw@gmail.com}
% \altaffiliation[Also at ]{Physics Department, Ben Gurion University, Beer Sheva, Israel}%Lines break automatically or can be forced with \\
\author{Eduardo Guendelman}%
 \email{guendel@bgu.ac.il}
\affiliation{%
Physics Department, Ben Gurion University, Beer Sheva, Israel
}%

\date{\today}% It is always \today, today,

\begin{abstract}
As it is well known, one can lower the energy of the trivial perturbation QCD vacuum by introducing a non-vanishing chromomagnetic field strength. This happens because radiative corrections produce an effective action of the form $f(F_{\mu\nu}^aF^{\mu\nu a})$ with $f'(y_0)=0$ for some $y_0\neq 0$. However, a vacuum with a non zero field strength is not consistent with Poincare Invariance (PI). Generalizing this type of effective action by introducing, in the simplest way, a four index field strength $\partial_{[\mu}A_{\nu\alpha\beta]}$, which can have an expectation value without violating PI, we are lead to an effective action that can describe both a confinement phase and a perturbative phase of the theory. In the unconfined phase, the 4-index field strength does not introduce new degrees of freedom, while in the confined phase both 4-index field strength and ordinary gauge fields are not true degrees of freedom. The matching of these phases through membranes that couple minimally to the 3-index potentials from which the 4-index field strength derive, leads automatically to the MIT bag boundary conditions for the gauge fields living inside the bubble containing the perturbative phase.  
\end{abstract}

\keywords{QCD,  Confining phase, Perturbative phase, MIT boundary conditions, Four index field}

\maketitle

\section{Introduction}         
The fact that some condensates of gauge field wont be present in the vacuum of QCD in $3+1$ dimensions, has been argued from the well known instability of the perturbative QCD vacuum studied in Ref. \cite{intro1}. There, at the one loop level, it was found that a condensate of the form $\langle A_{i a}\rangle=\frac{B}{2}\epsilon_{ijk}x_j \delta_{a 3}$, for example in the case of an $SU(2)$ gauge theory, can have an energy lower than that of the perturbative vacuum. In fact the one loop contribution to the vacuum energy of a state with a chromomagnetic field is 

\begin{equation}
E_{vac}=\dfrac{1}{2}B^2\left[ 1+bg^2 Ln\left( \dfrac{B}{\mu}\right)\right]
\label{eq:Evac}
\end{equation} 
where $b=\frac{11 C_A}{48 \pi^2}$, $C_A$ being the adjoint Casimir operator ($= N$ for $SU(N)$). $E_{vac}$ given by (\ref{eq:Evac}) has a minimum below zero at a non zero value of $B$, leading to the existence of a condensate. Once the existence of a condensate of gauge field is accepted, some problems are, of course, evident, since an explicit expectation value of $\langle A_{ia}\rangle$ breaks Lorentz invariance for example. Attempts to restore translation and rotation invariance were made in Refs. \cite{intro11,intro12,intro13,intro14,intro15,intro16,
intro17,intro18,intro19}. In \cite{intro110}, the rotation and homogeneity has been addressed  but still a full Poincare invariance formulation seems hard to obtain this way.    

Clearly, thinking about the non-stability of a gauge field condensate with the reasonable requirement of Poincare symmetry of the physical vacuum will be a nontrivial task, to say the least. One can think of different pictures where this is achieved. For example, in Ref. \cite{intro2}, in the case of QCD in $2+1$ dimensions, Poincare invariance is restored, while condensate are allowed in the vacuum but these all possible vector potentials, with all possible directions in color and real space are integrated with equal strengths. Such random background of coloured field is argued to lead to confinement of static charges \cite{intro3}, for other types of background gauge fields, color confinement effects have also been found \cite{intro4}.     

The above mentioned pictures where the conflict between the existence of a gauge condensate and Lorentz invariance use in a fundamental way at least one quantum mechanical concept, that is that of averaging in order to achieve an effectively Poincare invariant situation. Those effects can not be represented by a classical effective action. There is, however, a known classical gauge field \emph{action} which can acquire an expectation value without violating Poincare invariance. This is the case of a four index field strength $F_{\mu\nu\alpha\beta}$, which derives from a 3-index potential according to
\begin{equation}
F_{\mu\nu\alpha\beta}=\partial_{[ \mu}A_{\nu\alpha\beta]}
\end{equation}              
Such field strength $F_{\mu\nu\alpha\beta}$ is invariant under the transformation 
\begin{equation}
A_{\nu\alpha\beta}\rightarrow A_{\nu\alpha\beta}+\partial_{\nu [}V_{\alpha\beta ]}
\end{equation}
In 4-dimensional flat spacetime, it is an identity that 
\begin{equation}
F_{\mu\nu\alpha\beta}=\varphi \varepsilon_{\mu\nu\alpha\beta}
\end{equation}
$\varepsilon_{\mu\nu\alpha\beta}=0$ if any index is repeated and $\varepsilon_{\mu\nu\alpha\beta}$ is totally anti-symmetric. $\langle \varphi\rangle$ can be non zero and this is not in conflict with Poincare invariance of the vacuum.
%%%%%%%%%%%%%%%%%%%%%%%%%%%%%%%%%%%%%%%%%%%%%%%%%%%%%%%%%%%%%%%%%%%%%%%%%%%%%%%%%%%%%%%%%%%%%%%%%%%%%%%%%%%%%%%%%%%%%%%%%%%%%%%%%%%%%%%%%%%%%%%%%%%%%%%%%%%%%%%%%%

\section{4-index field strength and phases of the theory: Confining and Perturbative}

We consider an effective action that depends on the contributions 
\begin{equation}
y=F_{\mu\nu}^a F^{\mu\nu a}+\varepsilon^{\mu\nu\alpha\beta}\partial_{\mu}A_{\nu\alpha\beta}
\end{equation}
and where the contribution of the gauge fields to the Lagrangian density is of the from \footnote{
We can also include a the coupling of gauge fields to fermions (quarks), this will introduce another term in the action (\ref{eq:PerCon}), which will be added to $f(y)$, of the form 
\begin{equation}
\sum_{i=1}^3\overline{\Psi}_{i}\left( \centernot{\partial}-i g\centernot{A}_a\tau_{a}-m_{i}\right)\Psi_{i}
\nonumber
\end{equation}} 
\begin{equation}
S=\int f(y)d^{4}x \qquad  F_{\mu\nu}^{a}=\partial_{\mu}A_{\nu}-\partial_{\nu}A_{\mu}-gf^{a b c}A_{\mu}^{b}A_{\nu}^{c}
\label{eq:PerCon}
\end{equation}

%In such a case one must consider the following action  
It is straightforward to see that a variation of this action with respect to three form $A_{\nu\alpha\beta}$ will bring about the following equation
\begin{equation}
\varepsilon^{\mu\nu\alpha\beta}\partial _{\mu}\left(\dfrac{\partial f(y)}{\partial y}\right)=0
\end{equation} 
this gives 
\begin{equation}
\dfrac{\partial f(y)}{\partial y}=\omega=\text{Constant}
\end{equation}
%That gives $\dfrac{\partial f(y)}{\partial y}=\omega=\text{Constant}$, 
%this means that 
Meaning, for a truly non-linear function $f(y)$ (we exclude here therefore the case $f(y)=ay+b$, $a$ and $b$ being constants, for which $\frac{\partial f(y)}{\partial y}=a$ without imposing any conditions on $y$)    
\begin{equation}
y=F_{\mu\nu}^a F^{\mu\nu a}+\varepsilon^{\mu\nu\alpha\beta}\partial_{\mu}A_{\nu\alpha\beta}=\text{Constant}
\label{eq:y}
\end{equation} 
therefore $A_{\nu\alpha\beta}$ does not represent new degrees of freedom, since the field strength $F_{\mu\nu\alpha\beta}=\partial_{[\mu}A_{\nu\alpha\beta]}=\varphi\epsilon_{\mu\nu\alpha\beta}$, which represent the only physical gauge invariant object under the gauge transformations (3) can be solved from (\ref{eq:y}) in terms of $F_{\mu\nu}^{a}F^{\mu\nu a}$. 

The variation (\ref{eq:PerCon}) with respect to the gauge field $A_{\mu a}$ gives the following equation of motion
\begin{equation}
\partial_{\mu}\left(\dfrac{\partial f(y)}{\partial y}F^{\mu\nu a} \right)-g\dfrac{\partial f(y)}{\partial y}f^{a b c}A_{\mu}^{b}F^{\mu\nu c}=0
\label{eq:phases}
\end{equation}
If $\omega \neq 0$, i.e. we are not in the lower energy phase of the theory (the true vacuum), In this phase, the equations of motion are indistinguishable from those of classical QCD and they define a perturbative phase.

If we turn our attention to the true vacuum state ($\omega =0$), we see that a drastically different situation arises. In the true vacuum state $y=y_0$ and $\frac{\partial f(y)}{\partial y}\vert_{y=y_0}=0$, the gauge field equations of motion disappears and there is no meaningful equation for the gauge fields in this phase, except for (\ref{eq:y}), which can be solved for any gauge field configurations by appropriately choosing $A_{\nu\alpha\beta}$. That means that in this phase any gauge field configuration is allowed and that all gauge field configurations have constant energy density $f(y_0)$ \footnote{If  fermionic fields are also considered, one finds that the fermionic current, $j^{\nu a}=-\frac{1}{4} g\sum_{i=1}^3\overline{\Psi}_{i}\gamma^{\nu}\tau_{a}\Psi_{i}$, identically vanishes since the term representing the gauge fields contribution to the equation of motion disappears in this phase}.

Furthermore, the canonical momenta associated to $A_{\mu}^a$ all vanishes in the phase $\frac{\partial f(y)}{\partial y}\vert_{y=y_0}=0$
\begin{equation}
\pi_{\mu}^a=\dfrac{\partial L}{\partial \dot{A}_{\mu}^a}=0
\label{eq:constarint}
\end{equation}     
The vanishing of all the canonical conjugate momenta in the true vacuum phase means that, in this phase, the gauge fields are not true dynamical variables. In fact according to the Dirac constraint Hamiltonian theory, the first order constraints (\ref{eq:constarint}) generate the gauge invariance, generated by 
\begin{equation}
Q=\int \Lambda_a^{\mu}\pi_{\mu a}d^3x
%\ee  
\quad , \quad
%\be 
\delta A^{\mu}_a=[A_a^{\mu},Q]=\Lambda_a^{\mu}
\end{equation}
That is $A^{\mu}_a$ can be transformed into anything we want.
 
This is in agreement with ideas concerning the confinement phase of QCD in Ref. \cite{Polyakov}, where it was argued that the gauge invariance in the confinement phase is bigger then in the non-confinement phase. In Polyakov's language, we must have ``non-Abelian third kind gauge invariance" (chapter 5 in Ref. \cite{Polyakov}).   
%%%%%%%%%%%%%%%%%%%%%%%%%%%%%%%%%%%%%%%%%%%%%%%%%%%%%%%%%%%%%%%%%%%%%%%%%%%%%%%%%%%%%%%%%%%%%%%%%%%%%%%%%%%%%%%%%%%%%%%%%%%%%%%%%%%%%%%%%%%%%%%%%%%%%%%%%%%%%%%%%

\section{Dynamical Coupling in the Perturbative Vacuum}
We would like to show that equation (\ref{eq:phases}) in either phases with constant $\omega$, including the current $j^{\mu a}$, can be written in the original Yang-Mills form by rescaling the gauge fields $A_{\mu}^a$ and defining a new coupling constant in the following way   
\begin{equation}
A_{\mu}^a \rightarrow \tilde{A}_{\mu}^a=\omega^{\frac{1}{2}} A_{\mu}^a \quad , \quad g \rightarrow \tilde{g}=\dfrac{g}{\sqrt{\omega}}
\end{equation}
with those definitions the field strength tensor rescales to
\begin{equation}
 \tilde{F}_{\mu\nu}^{a}=\partial_{\mu}\tilde{A}_{\nu}-\partial_{\nu}\tilde{A}_{\mu}-\tilde{g}f^{a b c}\tilde{A}_{\mu}^{b}\tilde{A}_{\nu}^{c}
\end{equation}
and equation (\ref{eq:phases}), including the fermionic fields, reads
\begin{equation}
\partial_{\mu}\tilde{F}^{\mu\nu a}-\tilde{g} f^{a b c}\tilde{A}_{\mu}^{b}\tilde{F}^{\mu\nu c}=-\dfrac{1}{4} \tilde{g}\sum_{i=1}^3\overline{\Psi}_{i}\gamma^{\nu}\tau_{a}\Psi_{i}
\end{equation}
the effective coupling constant that results depends on the vacuum though $\omega$. As one approaches the true vacuum phase, where $\omega$ is equal to zero, we see that the effective coupling constant approaches infinity this is consistent with the idea that such a vacuum represent a confining phase of the theory.

\section{Bag Formation and Derivation of The MIT Boundary Conditions}
We now address the question of the possible coexistence of these two phases, that is whether is it possible for example to form bags of perturbative phase in the middle of a true vacuum state, the confinement phase. As we will see, the answer to this question is positive, such an effect can be produced if the minimal coupling of the $3$-index gauge potential $A_{\nu\alpha\beta}$ to $2+1$ membranes is considered \cite{1}, that is, we add to the action (\ref{eq:PerCon}) a term of the form
\begin{equation}
S_{int}=\lambda\int A_{\nu\alpha\beta}\dfrac{\partial z^{\nu}}{\partial\xi^a}\dfrac{\partial z^{\alpha}}{\partial\xi^b}\dfrac{\partial z^{\beta}}{\partial\xi^c}\varepsilon^{abc}d^3\xi
\end{equation} 
Where $z(\xi)$ represents the location of the $2+1$ dimensional membrane in the four dimensional spacetime and where $\xi^a$ and $\varepsilon^{abc}$ are, respectively, the coordinates defined on the membrane and the totally anti-symmetric 3-dimensional Levi-Civita tensor.

Here we present the study of the theory (\ref{eq:PerCon}) neglecting gravitational effects, having in mind a model for the strong interaction alone and in this context effects not considered in Ref.  \cite{1}. We can include a coupling of gauge fields to fermions (quarks). In such a case one must consider the following action            
%\begin{equation}
%\begin{split}
%S&=\int \left[f(y)+\sum_{i=1}^3\overline{\Psi}_{i}\left(i \gamma^{\mu}D_{\mu}-m_{i} \right)\Psi_{i}\right]d^{4}x 
%\\
%&+\int\delta^4(x-x(\xi))A_{\nu\alpha\beta}\dfrac{\partial x^{\nu}}{\partial\xi^a}\dfrac{\partial x^{\alpha}}{\partial\xi^b}\dfrac{\partial x^{\beta}}{\partial\xi^c}\varepsilon^{abc}d^3\xi d^4x 
%\end{split}
% \label{eq:Action}
%\end{equation}

\begin{align}
 S=&\int \left[f(y)+\sum_{i=1}^3\overline{\Psi}_{i}\left(i \gamma^{\mu}D_{\mu}-m_{i} \right)\Psi_{i}\right]d^{4}x \nonumber \\
 &+\int\delta^4(x-x(\xi))A_{\nu\alpha\beta}\dfrac{\partial x^{\nu}}{\partial\xi^a}\dfrac{\partial x^{\alpha}}{\partial\xi^b}\dfrac{\partial x^{\beta}}{\partial\xi^c}\varepsilon^{abc}d^3\xi d^4x 
 \label{eq:Action}
\end{align}

\subsection{Three-form field variation}
We start by performing a variation of the action (\ref{eq:Action}) with respect to the three-form $A_{\nu\alpha\beta}$, this variation leads to the following expressions which we will denote by $I_1$, $I_2$ % is given by
\begin{equation}
I_1=\int \partial_{\mu}\left(\varepsilon^{\mu\nu\alpha\beta}\dfrac{\partial f(y)}{\partial y} \delta A_{\nu\alpha\beta}\right)d^4x
\label{I1}
\end{equation} 
and
\begin{align}
I_2=&\int -\varepsilon^{\mu\nu\alpha\beta}\partial _{\mu}\left(\dfrac{\partial f(y)}{\partial y}\right)d^4x \nonumber \\
&+\lambda \int\delta^4(x-x(\xi))\dfrac{\partial x^{\nu}}{\partial\xi^a}\dfrac{\partial x^{\alpha}}{\partial\xi^b}\dfrac{\partial x^{\beta}}{\partial\xi^c}\varepsilon^{abc}d^3\xi\delta A_{\nu\alpha\beta}d^4x
\label{I2}
\end{align} 
Turning our attention to the second part of this variation, $I_2$, one has the following equation 
\begin{equation}
\varepsilon^{\mu\nu\alpha\beta}\partial_{\mu}\left( \dfrac{\partial f(y)}{\partial y}\right) =\lambda \int\delta^4(x-x(\xi))\dfrac{\partial x^{\nu}}{\partial\xi^a}\dfrac{\partial x^{\alpha}}{\partial\xi^b}\dfrac{\partial x^{\beta}}{\partial\xi^c}\varepsilon^{abc}d^3\xi
\label{eq:A variation}
\end{equation} 
We are interested in obtaining spherically symmetric solutions, in which the membrane forms a closed surface, dividing space into inside and outside regions. Then the normal to the membrane, pointing from the inside to the outside, is given by \cite{2,3}:
\begin{equation}
\dfrac{1}{\sqrt{-\gamma}}\dfrac{\partial x^{\nu}}{\partial\xi^a}\dfrac{\partial x^{\alpha}}{\partial\xi^b}\dfrac{\partial x^{\beta}}{\partial\xi^c}\varepsilon^{abc}\varepsilon_{\mu\nu\alpha\beta}=-\epsilon n_{\mu} 
\label{eq:Normal}
\end{equation}  
%\footnote{the normal is defined as $\dfrac{\sqrt{g}}{\sqrt{^dg}} \dfrac{\partial x^{\nu}}{\partial\xi^a}\dfrac{\partial x^{\alpha}}{\partial\xi^b}\dfrac{\partial x^{\beta}}{\partial\xi^c}\varepsilon^{abc}\varepsilon_{\mu\nu\alpha\beta}$, where $d=D-1$, in flat space $\sqrt{-g}=1$ \cite{2}--\cite{3},this normal points from the inside to the outside and $\epsilon=\pm 1$.}
Multiplying (\ref{eq:A variation}) by $n^{\sigma}$ and $\varepsilon_{\sigma\nu\alpha\beta}$ and using (\ref{eq:Normal}), we see that equation (\ref{eq:A variation}) shows that on either side of the membrane $\frac{\partial f(y)}{\partial y}$ is constant, and that these two values of $\frac{\partial f(y)}{\partial y}$ differ in magnitude $\vert\lambda\vert$. We will choose $\frac{\partial f(y)}{\partial y}$ to be zero outside the membrane (the outer region is the lower energy phase of the theory, the true vacuum) and have a non-vanishing value inside, in this manner we bring about the two different phases of the theory, the confining and perturbative phases.

For the first part of this variation, $I_1$, %which can be written as 
%\be
%\oint_{S} \varepsilon^{\m\n\a\a}\dfrac{\partial f(y)}{\partial y}n_{\m} \d A_{\n\a\b}ds
%\ee 
we notice that since one can not use (\ref{eq:A variation}), this integral will vanish from pure variational principles arguments, that is, the "end" points are held fixed, so $\delta A_{\nu\alpha\beta}$ is zero at the ends.

%%%%%%%%%%%%%%%%%%%%%%%%%%%%%%%%%%%%%%%%%%%%%%%%%%%%%%%%%%%%%%%%%%%%%%%%%%%%%%%%%%%%%%%%%%%%%%%%%%%%%%%%%%%%%%%%%%%%%%%%%%%%%%%%%%%%%%%%%%%%%%%%%%%%%%%%%%%%%%%
\subsection{Gauge Field Variation}
We now torn to the variation of the action (\ref{eq:Action}) with respect to $A_{\nu}^{a}$, we again denote the different parts by $I_3$ and $I_4$ given by
\begin{equation}
I_3=\int \partial_{\mu}\left(\dfrac{\partial f(y)}{\partial y}F^{\mu\nu a} \delta A^{a}_{\nu}\right)d^4x
\label{I3}
\end{equation}
and
\begin{align}
I_4=&\int\left(-\partial_{\mu}\left(\dfrac{\partial f(y)}{\partial y}F^{\mu\nu a} \right)-g\dfrac{\partial f(y)}{\partial y}f^{a b c}A_{\mu}^{b}F^{\mu\nu c}\right)d^4x \nonumber \\
&-\int\dfrac{i}{4} g\sum_{i=1}^3\overline{\Psi}_{i}\gamma^{\nu}\tau_{a}\Psi_{i}\delta A^{a}_{\nu}d^4x
\label{I4}
\end{align}
The second part of this variation, $I_4$, will give us the gauge field equations of motion
\begin{align}
&\partial_{\mu}\left(\dfrac{\partial f(y)}{\partial y} F^{\mu\nu a}\right)+gf^{a b c}A_{\mu}^{b}\dfrac{\partial f(y)}{\partial y}F^{\mu\nu c}= j^{\nu a} \nonumber \\
&=-\dfrac{i}{4} g\sum_{i=1}^3\overline{\Psi}_{i}\gamma^{\nu}\tau_{a}\Psi_{i} 
\label{FieldV}  
\end{align}
%\begin{equation}
%\partial_{\mu}\left(\dfrac{\partial f(y)}{\partial y} F^{\mu\nu a}\right)+gf^{a b c}A_{\mu}^{b}\dfrac{\partial f(y)}{\partial y}F^{\mu\nu c}=-\dfrac{i}{4} g\sum_{i=1}^3\overline{\Psi}_{i}\gamma^{\nu}\tau_{a}\Psi_{i} \equiv j^{\nu a}
%\label{FieldV}  
%\end{equation}
This equation, when expanded with the use of (\ref{eq:A variation}) and the discussion that followed it, contains elements multiplied by a delta-function and a step-function. 
\begin{align}
&\lambda \left(\int\delta^4(x-x(\xi))\dfrac{\partial x^{\nu}}{\partial\xi^a}\dfrac{\partial x^{\alpha}}{\partial\xi^b}\dfrac{\partial x^{\beta}}{\partial\xi^c}\varepsilon^{abc}\varepsilon_{\mu\nu\alpha\beta}d^3\xi \right)F^{\mu\nu a} \nonumber \\
&+\dfrac{\partial f(y)}{\partial y}\left( \partial_{\mu}F^{\mu\nu a}+gf^{a b c}A_{\mu}^{b}F^{\mu\nu c}\right)= j^{\nu a}
\end{align} 
Where the first part is actually a delta-function (on the surface) times the normal (by using (\ref{eq:Normal})) multiplied by the field strength tensor, and the function $\frac{\partial f(y)}{\partial y}$ is a generalized step function. Following this naive approach, one could conclude that this contribution, containing a delta-function, must vanish by itself. This will lead to the conclusion that $n_{\mu}F^{\mu\nu a}=0$ on the surface, thus, deriving the well known MIT boundary conditions. 

We will follow now a more rigorous formalism, i.e., the variational principle, to derive this results. 
To do this we first notice that the $I_3$ part of the variation, given in (\ref{I3}), contains the generalized step-function $\frac{\partial f(y)}{\partial y}$, where now the use of (\ref{eq:A variation}) is allowed. The appearance of $\frac{\partial f(y)}{\partial y}=0$ (outside the membrane) "cuts-off" the outer region of the integral, i.e., the "outer end point". Furthermore, since there is a discontinuity in $\frac{\partial f(y)}{\partial y}$, and its "on-surface value" is not well defined, it can take any value from zero to $\lambda$ so we can not say anything about $\frac{\partial f(y)}{\partial y}$ and $\delta A_{\nu}^a$ at the surface. 

Following these arguments, we conclude that the vanishing of this integral, (\ref{I3}), which can be written as  
\begin{equation}
\oint_{S}\dfrac{\partial f(y)}{\partial y}F^{\mu\nu a}n_{\mu} \delta A^{a}_{\nu}ds
\end{equation}
must be due to the boundary condition $n_{\mu}F^{\mu\nu a}=0$ on the surface, and  \textbf{not} because the potential $A^{a}_{\nu}$ is held fixed on the membrane. We then have a derivation of the famous and well known MIT boundary conditions from an action and variational principles. One should also note that there is no current in the confining phase, meaning the fermion fields $\Psi_i(x)$ are associated with the inner part, the perturbative phase, and $n_{\mu}j^{\mu a}=0$ on the surface as well.

%%%%%%%%%%%%%%%%%%%%%%%%%%%%%%%%%%%%%%%%%%%%%%%%%%%%%%%%%%%%%%%%%%%%%%%%%%%%%%%%%%%%%%%%%%%%%%%%%%%%%%%%%%%%%%%%%%%%%%%%
\section{Application to TMT}
Two Measure Theory (TMT) is a new class of gravity theories based on the idea that the action integral may contain a new metric-independent measure of integration. For example, in D = 4 space-time dimensions the new measure density can be built out of four auxiliary scalar fields $\varphi^i$ ($i=1,2,3,4$):
\begin{equation}
\Phi(\varphi)=\dfrac{1}{4!}\varepsilon^{\mu\nu\alpha\beta} \varepsilon_{ijkl} \partial_{\mu} \varphi^i \partial_{\nu} \varphi^j \partial_{\alpha} \varphi^k \partial_{\beta} \varphi^l
\end{equation} 
An action which incorporate this new measure, and the previous idea will take the following form 
\begin{equation}
S=\int_{\mathcal{M}} \Phi f(y) d^4x +\lambda \int_{\Sigma} A_{\nu\alpha\beta}dx^{\nu}\wedge dx^{\alpha}\wedge dx^{\beta}
\end{equation}
\begin{equation}
\quad y=\mathcal{L}_{g}(g_{\mu\nu},R_{\mu\nu}(\Gamma))+\mathcal{L}_{m}+\dfrac{\varepsilon ^{\mu\nu\alpha\beta}\partial_{\mu}A_{\nu\alpha\beta}}{\sqrt{-g}}
\end{equation}
We will relax the Levi-Civita condition on the metric and obtain the field equations following the Palatini approach, in which metric and connection are regarded as two physically independent entities \cite{palatini}. This implies that both metric and connection must be determined by solving the irrespective equations obtained through the application of the variational principle.

Our first step will be to take the variation with respect to $\varphi^{i}$ of the new measure field $\Phi$, this result of this variation is given by 
\begin{equation}
\partial_{\mu}f(y)=0\rightarrow \dfrac{\partial f(y)}{\partial y}\dfrac{\partial y}{\partial x^{\mu}}=0
\end{equation} 
The non-trivial solution of this equation is when $\frac{\partial y}{\partial x^{\mu}}=0$, which implies $ y=const$, this also implies $\frac{\partial f(y)}{\partial y}=const$, therefore, there is no jump in $\frac{\partial f(y)}{\partial y}$ across the boundary, in contrast to the previous section. Here the discontinuity will manifest in a new parameter $\chi$, which we will present next by taking the variation with respect to the three-form $A_{\nu\alpha\beta}$, this part of the variation is very much like the previous one, only this time the jump across the boundary is in the ratio of the two measures\footnote{ the jump is in the ratio of the two measures $\frac{\Phi}{\sqrt{-g}}$ since we have already concluded that $\frac{\partial f(y)}{\partial y}$ is constant }
\begin{equation}
\chi\equiv \dfrac{\Phi}{\sqrt{-g}}\dfrac{\partial f(y)}{\partial y}
\end{equation}
which means that $\chi$ is a generalized step function.

For the gravitational part of the action, the $g^{\mu\nu}$ variation will contribute the following equation
\begin{equation}
%\Phi \dfrac{\partial f(y)}{\partial y}
\chi\left(\dfrac{\partial(\mathcal{L}_{g}+\mathcal{L}_{m})}{\partial g^{\mu\nu}}+\dfrac{g_{\mu\nu}\varepsilon ^{\rho\sigma\alpha\beta}\partial_{\rho}A_{\sigma\alpha\beta}}{2\sqrt{-g}} \right)=0
\label{eq:g_variation}
\end{equation} 
We have seen that $y$ is constant, call it $y_{0}$, we can then use this to eliminate the 3-form field contribution 
\begin{equation}
\dfrac{g_{\mu\nu}\varepsilon ^{\rho\sigma\alpha\beta}\partial_{\rho}A_{\sigma\alpha\beta}}{2\sqrt{-g}}
\end{equation} 
by substituting it with $y_{0}-\mathcal{L}(g_{\mu\nu},R_{\mu\nu}(\Gamma))-\mathcal{L}_{matter}$ in (\ref{eq:g_variation}), this give rise to
\begin{equation}
%\Phi \dfrac{\partial f(y)}{\partial y}
\chi\left(\dfrac{\partial(\mathcal{L}_{g}+\mathcal{L}_{m})}{\partial g^{\mu\nu}}-\dfrac{\mathcal{L}_{g}+\mathcal{L}_{m}}{2}g_{\mu\nu}+\dfrac{y_0}{2}g_{\mu\nu} \right)=0
\label{eq:EOM1}
\end{equation}
Since $\chi$ is a generalized step function we conclude that equation (\ref{eq:EOM1}), assuming $\chi\neq0$ on either side of the membrane, actually describes two space-time regions, separated by a thin shell, and these two regions contain a dynamically generated cosmological constant.

%%%%%%%%%%%%%%%%%%%%%%%%%%%%%%%%%%%%%%%%%%%%%%%%%%%%%%%%%%%%%%%%%%%%%%%%%%%%%%%%%%%%%%%%%%%%%%%%%%%%%%%%%%%%%%%%%%%%%%%%

\section{Summary and Future work}

We have seen that an effective action with a four index field strength $\partial_{[\mu}A_{\nu\alpha\beta]}$ can describe both a confinement and perturbative phases. In the unconfined phase, the 4-index field strength does not introduce new degrees of freedom, while in the confined phase both 4-index field strength and ordinary gauge fields are not true degrees of freedom. Matching the two phases through membranes that couple minimally to the 3-index potentials, from which the 4-index field strength derive, leads automatically to the MIT-bag boundary conditions, from a purely variational principles, and to a description of a ``confining" theory in flat space time. 
 
As an application to curved space we studied here one case, in the frame work of the ``Two Measure Theory", where we find that the coupling of the three index field, $A_{\mu\nu\alpha}$, to a $2+1$ membrane induces a dynamical generation of a cosmological constant. Another application is found in Ref. \cite{EduMah1}, where we consider a modified\footnote{ we replace the standard measure density $e = \sqrt{-g}$ by an alternative measure density $\Phi(B)\equiv \frac{1}{3!}\varepsilon^{\mu\nu\kappa\lambda} \partial_\mu B_{\nu\kappa\lambda} $ 
}
 supergravity action in $4-$ dimension supplemented with the field-strength of a 3-index antisymmetric tensor gauge field $H_{\mu\nu\alpha}$, the action is given in the following form 
\begin{equation}
S =\frac{1}{2\kappa^2} \int d^4 x \Phi(B)
\left[ R(\omega,e) - {\bar\psi}_\mu \gamma^{\mu\nu\lambda} D_\nu \psi_\lambda 
+ \frac{\varepsilon^{\mu\nu\kappa\lambda}}{3!e}\partial_\mu H_{\nu\kappa\lambda} \right]
\label{eq:mSG-action}
\end{equation}

we find that the role of $H_{\mu\nu\alpha}$ in the modified-measure action (\ref{eq:mSG-action}) is to absorb, under local supersymmetry transformation, the total derivative term coming from the transformation
\begin{equation}
\delta_{\epsilon} \Big( e \big[ R(\omega,e) 
- {\bar\psi}_\mu \gamma^{\mu\nu\lambda} D_\nu \psi_\lambda \big]\Big) = 
\partial_\mu \big[ e\big({\bar\varepsilon}\zeta^\mu\big)\big]
\label{eq:local-susy-L}
\end{equation}
 
In Ref. \cite{EduMah2} we have considered a theory describing confinement, in flat spacetime, through a non-linear Maxwell term in the form of 
 \begin{equation}
 L=-\dfrac{1}{4} F_{\mu\nu}F^{\mu\nu}-\dfrac{f}{2}\sqrt{-F_{\mu\nu}F^{\mu\nu}} \quad , \quad f>0
  \label{eq:Previous}
\end{equation}  
When we coupled this Lagrangian to gravity we find that a genuinely charged matter source of gravity and electromagnetism may appear electrically neutral to an external observer. The ``charge-hiding" effect occurs in a self-consistent wormhole solution which connects a non-compact ``universe", comprising the exterior region of Schwarzschild (anti-)de-Sitter (or purely Schwarzschild) black hole beyond the internal (Schwarzschild) horizon, to a Levi-Civita-Bertotti Robinson type (``tube-like") ``universe", with two compactified dimensions, via a wormhole ``throat" occupied by the charged timelike brane. In this solution the whole electric flux produced
by the charged timetlike brane is expelled into the compactified Levi-Civita-Bertotti-
Robinson-type ``universe" and, consequently, the brane is detected as neutral by an external observer. We would like to show that the ``hiding" effect is not restricted to Lagrangian of the form (\ref{eq:Previous}), meaning, one can find ``hiding" solutions also for (\ref{eq:PerCon}) coupled to Einstein gravity. %And by doing so, generalize, in the future, the result to all theories describing confinement.    

%%%%%%%%%%%%%%%%%%%%%%%%%%%%%%%%%%%%%%%%%%%%%%%%%%%%%%%%%%%%%%%%%%%%%%%%%%%%%%%%%%%%%%%%%%%%%%%%%%%%%%%%%%%%%%%%%%%%%%%%%%%%%%%%%%%%%%%%%%%%%%%%%%%%%%%%%%%%%%%%%%%%%%%%%%%%%%%%%%%%%%%%%%%%%%%%%%%%%%%%%%%%%%%%%%%%%%%%%%%%%%%%%%%%%%%%%%%%%%%%%%%%%%%%%%%%%%%%%%%%%%%%%%%%%%%%%%%%%%%%%%%%%%%%%%%%%%%%%%%%%%%%%%%%%%%%%%%%%%%%%%%%%%%%%%%%%%%%%%%%%%%%%%%%%%%%%%%%%%%%%%%%%%%%%%%%%%%%%%%%%%%%%%%%%%%%%%%%%%%%%%%%%%%%%%%%%%%%%%%%%%%%%%%%%%%%%%%%%%%%%%%%%%%%%%%%%%%%%%%%%%%%%%%%%%%%%%%%%%%%%%%

\end{document}